\documentclass[11pt]{article}

\usepackage{verbatim}
\usepackage{amsmath,amscd}
\usepackage{amsfonts}
\usepackage{amssymb}
\usepackage{bbm}
\usepackage{latexsym}
\usepackage{graphics}
\usepackage{lscape} 
\usepackage{epsfig}
\usepackage{color}

\usepackage{amscd}
\usepackage{cite}

\renewcommand{\d}{\mathrm{d}}

\newcommand{\captn}[1]{\vspace{-3ex}\caption{\small #1}}

\DeclareMathSymbol{\mg}{\mathrel}{symbols}{"1D}

\newcommand{\ga}{\alpha}
\newcommand{\gb}{\beta}
\renewcommand{\gg}{\gamma}
\newcommand{\gd}{\delta}

\newcommand{\gf}{\phi}
\newcommand{\gvf}{\varphi}

\newcommand{\gm}{\mu}

\newcommand{\gl}{\lambda}
\newcommand{\gr}{\rho}
\newcommand{\gth}{\theta}

\newcommand{\gs}{\sigma}
\newcommand{\gt}{\tau}
\newcommand{\go}{\omega}

\newcommand{\gp}{\pi}
\newcommand{\gps}{\psi}

\newcommand{\gG}{\Gamma}

\newcommand{\gF}{\Phi}

\newcommand{\gX}{\Xi}
\newcommand{\gL}{\Lambda}
\newcommand{\gS}{\Sigma}
\newcommand{\gTh}{\Theta}

\newcommand{\gPs}{\Psi}

\newcommand{\cA}{{\cal A}}

\newcommand{\cD}{{\cal D}}

\newcommand{\cF}{{\cal F}}

\newcommand{\cO}{{\cal O}}

\newcommand{\cR}{{\cal R}}

\newcommand{\cT}{{\cal T}}

\newcommand{\tD}{{\widetilde D}}

\newcommand{\Tr}{\mbox{Tr}}
\newcommand{\tr}{\text{tr}}

\newcommand{\lra}{\longrightarrow}

\newcommand{\ra}{\rightarrow}

\newcommand{\der}{\partial}

\newcommand{\dsp}{\displaystyle}

\newcommand{\Kh}{K\"{a}hler}
\newcommand{\beq}{\begin{equation}}
\newcommand{\eeq}{\end{equation}}
\newcommand{\barr}{\begin{array}}
\newcommand{\earr}{\end{array}}
\newcommand{\equ}[1]{\begin{gather} #1 \end{gather}}
\newcommand{\equa}[1]{\begin{align} #1 \end{align}}

\newcommand{\enums}[1]{\begin{enumerate} #1 \end{enumerate}}
\newcommand{\tabu}[2]{\begin{tabular}{#1} #2 \end{tabular}}
\newcommand{\arry}[2]{\begin{array}{#1} #2 \end{array}}

\newcounter{oldcounter}
\newenvironment{subeqns}{
\addtocounter{equation}{1}
\setcounter{oldcounter}{\value{equation}}
\setcounter{equation}{0}

}
{
\setcounter{equation}{\value{oldcounter}}
\vspace{-10pt}\\
}

\newcommand{\bder}{\bar\partial}
\newcommand{\bz}{{\bar z}}

\newcommand{\bY}{{\overline Y}}

\newcommand{\bgs}{{\bar\sigma}}

\newcommand{\Intr}{\mathbb{Z}}

\newcommand{\ba}[2]{\[\begin{array}{#2}\label{#1}}
\newcommand{\ea}{\end{array}\]}
\newcommand{\be}{\begin{equation}}
\newcommand{\ee}{\end{equation}}
\newcommand{\bea}{\begin{eqnarray}}
\newcommand{\eea}{\end{eqnarray}}

\begin{document}

\thispagestyle{empty}

\begin{flushright}
LMU-ASC 27/11 
\\
\end{flushright}
\vskip 2 cm
\begin{center}
{\Large {\bf Green--Schwarz Mechanism in Heterotic (2,0) Gauged Linear Sigma 
\\[1ex]
Models: Torsion and NS5 Branes} 
}
\\[0pt]

\bigskip
\bigskip {\large
{\bf Michael Blaszczyk$^{a,}$}\footnote{
E-mail: michael@th.physik.uni-bonn.de},
{\bf Stefan Groot Nibbelink$^{b,}$}\footnote{
E-mail: Groot.Nibbelink@physik.uni-muenchen.de},
{\bf Fabian Ruehle$^{a,}$}\footnote{
E-mail: ruehle@th.physik.uni-bonn.de}
\bigskip }\\[0pt]
\vspace{0.23cm}
${}^a$ {\it 
Bethe Center for Theoretical Physics and Physikalisches Institut der Universit\"at Bonn, \\
~~Nussallee 12, 53115 Bonn, Germany
}\\[1ex] 
${^b}$ {\it 
Arnold Sommerfeld Center for Theoretical Physics,\\
~~Ludwig-Maximilians-Universit\"at M\"unchen, 80333 M\"unchen, Germany
 \\} 
\bigskip
\end{center}

\subsection*{\centering Abstract}
Heterotic string compactifications can be conveniently described in the language of (2,0) gauged linear sigma models (GLSMs). Such models allow for Fayet--Iliopoulos (FI)--terms, which can be interpreted as K\"ahler parameters and axions on the target space geometry. We show that field dependent non--gauge invariant FI--terms lead to a Green--Schwarz--like mechanism on the worldsheet which can be used to cancel worldsheet anomalies. However, given that these FI--terms are constrained by quantization conditions due to worldsheet gauge instantons, the anomaly conditions turn out to be still rather constraining. Field dependent non--gauge invariant FI--terms result in non--K\"ahler, i.e.\ torsional, target spaces in general. When FI--terms involve logarithmic terms, the GLSM seems to describe the heterotic string in the presence of Neveu--Schwarz (NS)5 branes. In particular,  the GLSM leads to a decompactified target space geometry when anti--NS5 branes are present.

\newpage 
\setcounter{page}{1}
\setcounter{footnote}{0}
\section{Introduction and conclusions}


In this work we study heterotic (2,0) gauged linear sigma models (GLSMs) with field dependent Fayet--Iliopoulos (FI)--terms. Given that there are various motivations for this investigation, we begin the introduction by listing them. Then, after we have summarized the main advantages of using GLSMs to describe string compactifications, we state our main results.

String theories in general, and the heterotic string~\cite{Gross:1985fr,Gross:1985rr} in particular, are promising candidates for unifying descriptions of particle physics and gravitational phenomena. There has been huge progress in the recent years in identifying Minimal Supersymmetric Standard Model (MSSM) candidates within this framework. In the context of the type--II strings there has been a lot of progress in understanding flux vacua and the resulting mechanism of moduli stabilization. The understanding of flux configurations of the heterotic string is much harder to develop as it always involves non--trivial gauge bundles. Let us consider the various motivations in more detail: 

\subsubsection*{Heterotic orbifold and resolution model building}

One possibility of heterotic model building is the compactification on toroidal orbifolds, i.e.\ discrete quotients of six dimensional tori~\cite{Dixon:1985jw,Dixon:1986jc,Ibanez:1988pj}. In the so--called ``mini--landscape'' study on the $T^6/\Intr_\text{6--II}$ orbifold a large number of models has been identified, which upon switching on a certain number of vacuum expectation values (VEVs) lead to MSSM--like spectra~\cite{Lebedev:2006tr,Buchmuller:2006,Buchmuller:2006ik,Lebedev:2006kn,Lebedev:2007hv}. Hence the orbifold theory is driven away from the orbifold point resulting in (partially) smoothed out Calabi--Yau (CY) geometries. 

In order to understand the properties of the resulting CY and the gauge backgrounds it can support, one needs to systematically study the blowup process. In refs.~\cite{Douglas:1997de,Denef:2004dm,Lust:2006zh,Reffert:2007im} the topology of the smooth CY resulting from orbifold resolutions was discussed using toric geometry methods~\cite{Fulton,Hori:2003ic}. There has been recent progress in realizing Abelian gauge fluxes (line bundles) on both non-compact and compact resolutions~\cite{Honecker:2006qz,Nibbelink:2007pn,Nibbelink:2008tv,Nibbelink:2009sp}, which sometimes can even be constructed explicitly~\cite{Nibbelink:2007rd,GrootNibbelink:2007ew,Nibbelink:2008qf}. 

For the ``mini--landscape'' models these results imply that this blowup process can never lead to a completely smooth CY geometry since the hypercharge (or the weak $SU(2)$) gets broken~\cite{Nibbelink:2009sp}\footnote{Of course, these models are fine, as long as one is considering them as small perturbations of the orbifold theory.}. To avoid this  in full blowup another interesting MSSM realization was constructed on a $\Intr_2\times\Intr_2$ orbifold as an $SU(5)$ GUT that was subsequently broken down to the MSSM using a freely acting $\Intr_\text{2,free}$ involution~\cite{Blaszczyk:2009in}.  (For related $(\Intr_2)^n$ orbifolds see \cite{Donagi:2008xy}.)  In a follow--up paper~\cite{Blaszczyk:2010db} it was shown that this model can in principle be blown up completely without breaking the hypercharge.

\subsubsection*{Heterotic Calabi--Yau model building}

The choice of a toroidal orbifold is very restrictive. A more general approach is the compactification~\cite{Candelas:1985en} on smooth CY threefolds with stable vector bundles~\cite{Donaldson:1985,Uhlenbeck:1986}, as there are only few distinct orbifold geometries compared to the vast amount of known CY threefolds, see e.g.~\cite{Kreuzer:2000xy}.

In a CY compactification the  metric is Ricci flat at leading order in the $\alpha'$ expansion and would lead in any case to a conformal description with field dependent kinetic terms, i.e.\ a non--linear sigma model (NLSM). Hence, in order to be able to work with these theories one has to study the supergravity limit of heterotic string theory. This only allows for partial access to the low energy data of the theory. Seen as a perturbation theory in the string scale, the validity of the supergravity approach can only be guaranteed in the large volume limit.  However, even in this regime one can never be sure to have captured all stringy consistency conditions. One example was provided by the resolutions of the $T^6/\Intr_2\times\Intr_2\times\Intr_\text{2,free}$ mentioned above, where an additional modular invariance condition was identified for the $\Intr_\text{2,free}$ element, which is invisible using pure supergravity techniques\ \cite{Blaszczyk:2009in,Blaszczyk:2010db}.

As smooth CY spaces with stable bundles are very complicated to construct, finding the MSSM--like models has proven very difficult, especially because of the issue of bundle stability~\cite{Gomez:2005ii}. Ongoing efforts of refs.~\cite{Donagi:1999ez,Donagi:2000zs,Donagi:2000fw,Braun:2005ux,Braun:2005nv} have resulted in MSSM-like candidates, see \cite{Bouchard:2005ag,Bouchard:2008bg} and \cite{Anderson:2009mh,Anderson:2011ns}.

\subsubsection*{Heterotic flux vacua and torsion manifolds}

It has been realized in the type--II setting that flux vacua can effectively be used to fix a large number of moduli~\cite{Kachru:2002he,Kachru:2003aw,Denef:2004dm,Douglas:2006es,Denef:2007pq}. The development of flux vacua in the heterotic string context is much more involved due to the necessity of constructing stable vector bundles. Moreover, since such vacua have a non--trivial three form flux $H_3$, the manifold has torsion and is therefore no longer K\"ahler~\cite{Strominger:1986uh,Becker:2003yv,Becker:2003sh}. In light of this, it is not surprising that the first heterotic torsion geometries were constructed via a sequence of dualities from a type--IIB orientifold~\cite{Dasgupta:1999ss,Becker:2002sx,Becker:2009df}. These torsion geometries with consistent gauge backgrounds were studied in~\cite{Fu:2006vj,Becker:2006et,Fu:2008ga,Andreas:2010qh,Andreas:2010cv}. All these discussions of heterotic flux vacua were primarily performed using the effective target space supergravity description. The only concrete attempts to arrive at a microscopic heterotic string description has been undertaken in refs.~\cite{Adams:2006kb,Adams:2009av,Adams:2009zg}. Their observation that these torsion geometries can be described by non--invariant FI--terms has been the starting point for our investigation.

\subsubsection*{Microscopic description of heterotic strings in the presence of NS5 branes}

In most of the constructions of MSSMs on smooth CY--spaces the Bianchi identities are only satisfied in the presence of Neveu--Schwarz (NS)5 branes. In order to ensure that there are indeed NS5 branes (and not anti--NS5 branes which would break all supersymmetries~\cite{Becker:1995kb}), $\text{ch}_2(V) - \text{ch}_2(TX)$ has to be an effective class on the CY space $X$ with vector bundle $V$, see e.g.\ \cite{Aldazabal:1996du,Donagi:1998xe,Blumenhagen:2005zg,Honecker:2006dt,Anderson:2008uw}. On the worldsheet of the NS5 brane an enhanced gauge symmetry can arise, allowing gauge groups with rank larger than 16 to arise in the heterotic string as well~\cite{Witten:1995gx}. 

As these are heterotic solitons~\cite{Strominger:1990et}, non--perturbative effects come to rescue the consistency of these models. The heterotic string is not able to give a microscopic description of these NS5 branes. However, one might hope that it is possible to quantize the heterotic string in the presence of these heterotic solitons. In ref.\ \cite{Carlevaro:2009jx} this was attempted in the specific case of a resolved conifold. As we will discuss next, GLSMs provide a convenient framework for understanding the dynamics of heterotic strings. Therefore it would be very interesting to be able to describe NS5 branes in this context.

\subsection*{GLSM Methodology}

In order to avoid the complication of an NLSM description of smooth CY spaces, the requirement of conformality can be left aside for the time being and we can consider two dimensional GLSMs~\cite{Witten:1993yc}. These are supersymmetric theories on the worldsheet, containing an Abelian gauge sector, whose vacuum manifolds are naturally complete intersections of hypersurfaces in toric ambient spaces~\cite{Witten:1993yc,Distler:1987ee,Distler:1992gi,Distler:1995mi}. In addition, they provide tools to describe vector bundles in terms of monads. Since many known CYs are constructed this way, a GLSM can be considered a simpler approach to string theories described on them. Now, since in two dimensions the gauge couplings and some kinetic terms are of non--vanishing mass dimension, GLSMs are not conformal. Nevertheless, it is believed that in the infrared limit, where all dimensionful parameters are sent to infinity and massive modes are integrated out, the theory flows to a conformal NLSM~\cite{Silverstein:1994ih,Silverstein:1995re,Basu:2003bq,Beasley:2003fx}.

As discussed above, the choice of the vector bundle is crucial for the spectrum in the resulting four dimensional theory. The simplest choice is the standard embedding which automatically possesses $N=(2,2)$ supersymmetry. However, as the resulting four dimensional gauge group is $E_6$, one needs to construct more general vector bundles which reduce the worldsheet supersymmetry to $(2,0)$. Since in contrast to $(2,2)$ theories $(2,0)$ models are chiral, gauge anomalies can occur in the GLSM. Requiring absence of these anomalies puts severe constraints on the choice of the vector bundle~\cite{Distler:1995mi}. 

GLSMs might provide a framework which is able to smoothly interpolate between orbifold and smooth CY theories with vector bundles. Recently an explicit mapping between non--compact heterotic orbifold models and their resolutions with bundles was given~\cite{Nibbelink:2010wm}: The shifted momenta that characterize the twisted states, which take non--vanishing VEVs to generate the blowup, essentially define the gauge charges of a GLSM which might have a large volume interpretation as a CY space with a certain vector bundle. 
\\

\subsection*{Main results}

After this summary of the most important motivations for the present work, we are in the position to state our main findings: 
\enums{
\item[{\bf 1)}] {\bf Field dependent FI--term can cancel worldsheet anomalies} 
\\[1ex] 
We present a method to cancel gauge anomalies in $(2,0)$ GLSMs using a Green--Schwarz--like construction on the world--sheet. Following~\cite{Adams:2006kb} we introduce field--dependent Fayet--Iliopoulos terms. They are not gauge invariant precisely in such a way, that they can cancel the anomalous variation of the path integral measure. This weakening of the worldsheet gauge anomaly conditions results in more freedom in the construction of vector bundles for heterotic compactifications. The prime examples of such FI--terms we consider in this paper contain logarithmic terms of homogeneous and holomorphic polynomials of the coordinate fields. 
\item[{\bf 2)}] {\bf Field dependent FI--terms are subject to quantization conditions}  
\\[1ex] 
The coefficients in front of the field dependent FI--terms need to be quantized to ensure that the path integral is well--defined in the presence of worldsheet gauge instantons. These conditions can severely constrain the freedom to cancel worldsheet gauge anomalies. How strong these conditions are, depends on the gauge instantons that are allowed by the dynamics of the GLSM in question. 
\item[{\bf 3)}] {\bf Logarithmic FI--terms describe heterotic strings in the presence of NS5 branes} 
\\[1ex] 
Since GLSM FI--terms have the target space interpretation of the complexified K\"ahler moduli, their field dependence leads to a non--closed K\"ahler form. The resulting geometry is no longer K\"ahler and may be referred to as torsion geometry since there is a non--vanishing $H_3$ flux. Furthermore, the imaginary parts of logarithmic FI--terms lead to singular Kalb--Ramond background $B_2$, which results in an additional contribution $X_4 \sim \d(\d B_2)$ in the Bianchi identity $\d H_3 = X_4 + \tr \cR^2 - \tr \cF^2$. This contribution can be given the interpretation of NS5 branes wrapping certain internal curves which are  Poincar\'e dual to the class defined by $X_4$. 
\item[{\bf 4)}] {\bf Field dependent FI--terms may result in  decompactification}
\\[1ex] 
Field dependent FI--terms can lead to drastic modifications of the target space geometry and might even lead to decompactification. The reason is that the FI--parameters appear in the worldsheet D--term scalar potential controlling the target space geometry. In particular logarithmic terms appearing in the D-terms cause the resulting geometry to depart from being a standard symplectic quotient. Consequently, the divisor locus, corresponding to the polynomial of which the FI--term is the logarithm, is no longer part of the target space geometry of the string. Moreover, the resulting geometry may even become non--compact in the case of anti--NS5 branes. By an explicit example on the complete intersection Calabi--Yau (CICY) $\mathbbm{P}^7[2,2,2,2]$ we show that, the geometry in the presence of the NS5 branes can remain compact.  
}

\subsection*{Outlook}

In this work we have seen that the geometry can be decompactified  when the anomaly coefficients have the wrong sign. We interpret this as a signal that the logarithmic FI--terms on the worldsheet should be interpreted as anti--NS5 branes. An important check which is beyond the scope of the present paper is to show that target space supersymmetry is broken in this case.

\subsection*{Paper organization} 

We start with an overview of the basic ingredients of (2,0) heterotic GLSMs. In section \ref{sc:WGS} we describe how worldsheet gauge anomalies can be cancelled by field dependent FI--terms, and explain how stringent quantization conditions for them arise. Here we also argue that logarithmic FI--terms can be interpreted as describing NS5 branes. In the final section we study two examples: the well--known quintic and the CICY mentioned above.

\subsection*{Note added}

Part of this work was presented at the workshop ``Topological Heterotic Strings and (0,2) Mirror Symmetry'' at the ESI in Vienna. During that we became aware of the (at that time unpublished) work by Callum Quigley and Savdeep Sethi~\cite{Quigley:2011pv}, which also studies the potential consequences of logarithmic FI--terms in GLSMs.

\subsection*{Acknowledgments}

MB and FR thank the Ludwig Maximilians Universit\"at Munich (LMU) for hospitality. SGN would like to thank the Erwin Schr\"odinger International Institute for Mathematical Physics (ESI) in Vienna for kind hospitality. We would like to thank Allan Adams, Ralf Blumenhagen, Luca Carlevaro, Ron Donagi, James Gray, Leonhard Horstmeyer, Ilarion Melnikov, Adrian Mertens, Callum Quigley, Savdeep Sethi, Patrick Vaudrevange and Martijn Wijnholt for very useful discussions. This work was partially supported by the SFB-Tansregio TR33 ``The Dark Universe'' (Deutsche Forschungsgemeinschaft) and the European Union 7th network program ``Unification in the LHC era'' (PITN-GA-2009-237920). Furthermore, this work was supported by the LMUExcellent Programme.

\section{(2,0) gauged linear sigma models}

\begin{table}
\begin{center}
\renewcommand{\arraystretch}{1.2} 
\tabu{| c c || c | c | c c | c c c | }{
\hline 
\multicolumn{2}{|c||}{superfield} &   & gauge & \multicolumn{2}{|c|}{bosonic DOF} & \multicolumn{3}{|c|}{fermionic DOF}
\\
type & notation & dimension & charge & on & off & & on     & off  
\\ \hline\hline 
chiral & $\gPs^{a}$ & 0 & $(q_I)^a$ & $z^a$ & - & & $\gps^a$ & -
\\ 
chiral--Fermi & $\gL^{\ga}$ & 1/2 & $(Q_I)^\ga$ & - & $h^\ga$ & & $\gl^\ga$ & -  
\\ 
\hline 
gauge & $(V, A)^{I}$ & (0,1) & $0$ & $a_\gs^I, a_\bgs^I$ & $\tD^I$ & & $\gf^I$ & -  
\\ 
Fermi--gauge & $\gS^{i}$ & 1/2 & $0$ & $s^i$ & - & &$\gvf^i$ & -  
\\ \hline 
chiral & $\gF^m$ & 1 & $(q_I)^m$ & $x^m$ & - & & $\gps^m$ & -
\\
chiral--Fermi & $\gG^\gm$ & 3/2 & $(Q_I)^\gm$ & - & $h^\gm$ & & $\gg^\gm$ & - 
\\ \hline 
}
\renewcommand{\arraystretch}{1} 
\end{center} 
\captn{The superfield content of a gauged linear sigma model and their physical on- and off-shell degrees of freedom (DOF). 
\label{tb:Superfields}}
\end{table}

We start by introducing the field content of a generic (2,0) GLSM to set our notations and conventions. (For details on (2,0) GLSM see e.g.\ 
\cite{Distler:1992gi,Witten:1993yc,Distler:1995mi}.) The complete field content is summarized in Table~\ref{tb:Superfields}. A (2,0) GLSM contains a set of chiral superfields $\gPs^a$ and chiral--Fermi multiplets $\gL^\ga$. These superfields can be charged under bosonic gauge transformations 
\equ{
\gPs^a \ra e^{(q_I)^a \gTh^I}\, \gPs^a~, 
\qquad 
\gL^\ga \ra e^{(Q_I)^\ga \gTh^I}\, \gL^\ga~, 
\label{bos_gauging} 
}
with chiral superfield parameters $\gTh^I$ and gauge superfields $(V,A)^I$. For the gauge superfields we can write down a FI--term 
\equ{
W_\text{FI} = \frac 1{2\gp}\, \gr_J(\gPs) \, F^J~, 
\label{FI_term} 
}
where $F^J$ is the super gauge field strength such that $D_+ F^J| = -(\tD^J + i\, f_{\gs\bgs}^J)/2$ where $\tD^J$ is the auxiliary component of the gauge multiplet and the Abelian worldsheet gauge field strength is
$f_{\gs\bgs}^J = \der_\gs a^J_\bgs - \der_\bgs a^J_\gs$. (We use the notation: $\gs = (\gs^0 + \gs^1)/2$,  $\bgs = (\gs^0 - \gs^1)/2$ and 
$a_\gs = (a_0 + a_1)/2$,  $a_\bgs = (a_0 - a_1)/2$,  so that $\der = \der_0 + \der_1$ and $\bder = \der_0 - \bder_1$.) The parameters $b_J = \text{Re}(\gr_J)$ can be interpreted as \Kh\ parameters of the resulting target space geometry and the $\gb_J = \text{Im}(\gr_J)$ as the corresponding axions~\cite{Witten:1993yc}. Furthermore, a GLSM can be equipped with a number of fermionic gaugings 
\equ{
\gd_\gX \gL^\ga = M^\ga{}_{i}(\gPs) \, \gX^i~,  
}
with neutral chiral--Fermi parameters $\gX_i$ and Fermi--gauge superfields $\gS_i$.

In addition, the superfields $\gPs^a$ and $\gL^\ga$ may be subject to various holomorphic constraints
\equ{
P_\gm(\gPs) = 0~, 
\qquad 
N_{m \ga}(\gPs)\, \gL^\ga = 0~.  
\label{constraints} 
}
These constraints are encoded in the superpotentials 
\equ{
W_\text{geom} = \gG^\gm \, P_\gm(\gPs)~, 
\qquad 
W_\text{bundle} =  \gF^m\, N_{m \ga}(\gPs)\, \gL^\ga~.
}
Because of the mass dimensions specified in Table~\ref{tb:Superfields} the chiral--Fermi superfields $\gG^\gm$ and the chiral superfields $\gF^m$ can only appear linearly in the superpotential\footnote{The fact that these fields only appear linearly can equally well be enforced via their $U(1)_R$-charges.}. The holomorphic functions $P_\gm(\gPs)$ and $N_{m \ga}(\gPs)$ are subject to the requirement that the superpotential is gauge invariant under both bosonic and fermionic gaugings. This implies that in general also the $\gG^\gm$ transform under the fermionic gauge transformations 
\equ{
\gd_\gX \gG^\gm = \gF^m\, M_{mi}{}^\gm(\gPs)\, \gX^i~,
}
such that 
\equ{
  \gF^m \, M_{mi}{}^\gm(\gPs) \,P_\gm(\gPs)
+ \gF^m\, N_{m \ga}(\gPs)\, M^\ga{}_i(\gPs) = 0~. 
\label{cond_complex}
}

The first set of constraints in \eqref{constraints} can be interpreted geometrically as defining a number of hypersurfaces in the projective space spanned by the $\gPs^a$'s. Together with the fermionic gaugings that satisfy \eqref{cond_complex} the second set of constraints in \eqref{constraints} define a complex of vector bundles \cite{Witten:1993yc} 
\equ{
0 
\quad\lra\quad 
\cO^{N_\gS} 
\quad\stackrel{M}{\dsp \lra}\quad 
\bigoplus\limits_{\ga}^{} \cO_\ga 
\oplus
\bigoplus\limits_{\gm}^{} \cO_\gm
\quad\stackrel{N}{\dsp \lra}\quad  
\bigoplus\limits_{m}^{} \cO_m
\quad\lra\quad 
 0~,
}
where $N_\gS$ denotes the number of fermionic gaugings. This complex generically determines a holomorphic vector bundle via $V = \text{Ker}(N)/\text{Im}(M)$, called monad \cite{Blumenhagen:2011sq}.

\section{Worldsheet Green--Schwarz mechanism}
\label{sc:WGS}
\subsection{Gauge anomalies on the worldsheet}

The gauge symmetries of the GLSM with classical action $S_\text{cl}$ have to be preserved at the quantum level. Yet the effective Euclidean action 
\equ{
e^{-S_\text{eff}} = \int \cD \gPs \cD \gL \cD \gF \cD \gG\, 
e^{-S_\text{cl}}
}
might be gauge covariant rather than invariant because of worldsheet gauge anomalies. In two dimensions both bosons and fermions can induce anomalies in principle. However, since the coordinate fields consist of pairs of left-- and right--movers, only the fermions inside the chiral and the chiral--Fermi superfields may contribute to anomalies. They give contributions with opposite signs as their fermions are right-- and left--moving, respectively.

In detail gauge anomalies are characterized as follows. Denote by $\ga^I$ the scalar gauge parameters that are part of the lowest components of the chiral superfield gauge parameters $\gTh^I$. The structure of gauge anomalies $\gd_\ga S_\text{eff} = \int 2\gp i\, \go_{2,1}(\ga)$, is encoded via the descent equations~\cite{Zumino:1984rz,Alvarez:1984yi,Zumino:1985ws} 
\equ{
\d \go_3 = \go_4~, 
\qquad 
\gd_\ga \go_3 = d \go_{2,1}(\ga) 
\label{descent_equs}
}
by the second Chern character 
\equ{
\go_4 = \text{ch}_2(f_2) := \frac 12 \, \Tr \Big(\frac{f_2}{2\gp} \Big)^2
= \frac 12\, \sum_{I,J}\cA_{IJ} \, \frac{f_2^I}{2\gp} \, \frac{f_2^J}{2\gp}~, 
}
where $f_2^I$ denotes the field strength two--form $f_2^I = f_{\gs\bgs}^I \d\gs \d\bgs$ on the worldsheet. As we explain in detail below the descent equations do not determine the form of the mixed anomalies completely. The trace $\Tr$ over the full charged spectrum on the worldsheet determines the anomaly coefficients 
\equ{
\cA_{I J} :=  q_I \cdot q_J - Q_I \cdot Q_J ~,
\label{anom_coeffs}
}
in terms of the inner products 
\begin{subeqns} 
\equ{ 
Q_I \cdot Q_J := \sum_\ga (Q_I)^\ga (Q_J)^\ga + \sum_\gm (Q_I)^\gm (Q_J)^\gm~, 
\\[1ex] 
q_I \cdot q_J := \sum_a (q_I)^a (q_J)^a + \sum_m (q_I)^m (q_J)^m~,
}
\end{subeqns} 
that involve sums over all chiral and chiral--Fermi superfields, respectively.

The resulting anomalous variation of the effective action $S_\text{eff}$ 
\equ{
\gd_\gTh S_\text{eff} = 
\int \d^2 \gs \d \gth^+\,  W_\text{anom}(\gTh) + \text{h.c.}~, 
\qquad 
W_\text{anom}(\gTh) = 
\frac 1{4\gp}\, \sum_{I,J} \cA_{IJ} \, \gTh^I F^J~, 
\label{W_anom} 
}
can be encoded in an anomalous superpotential $W_\text{anom}$.

\subsubsection*{Shifting mixed anomalies around}

As mentioned above, even though the pure anomalies ($I=J$) are uniquely determined by the descent equations \eqref{descent_equs}, this is not the case for the mixed anomalies ($I\neq J$). The crucial point is that for mixed $U(1)$ anomalies the Chern--Simons three--form 
\equ{
\go_3^\text{mix} = 
\frac 1{(2\gp)^2}\,
\sum_{I< J} \cA_{IJ} \, 
\Big[ 
\big(1-c_{IJ}\big)\, a_1^I f_2^J + c_{IJ}\, a_1^J f_2^I 
\Big]~, 
}
and consequently, the anomalies 
\equ{ 
\go^\text{mix}_{2,1}(\ga) = 
\frac 1{(2\gp)^2}\, 
\sum_{I< J} \cA_{IJ} \, 
\Big[ 
\big(1- c_{IJ}\big)\, \ga^I f_2^J + c_{IJ}\, \ga^J f_2^I 
\Big]~, 
\label{anomalies}
}
are determined only up to some unknown constants $c_{IJ}$. Here we have assumed that we have specified some ordering, denoted by $I<J$, of the gauge indices. The appearance of these undetermined constants can be traced back to the regularization dependence in the computation of the gauge anomalies. Moreover, by using counter terms proportional to $\int a_1^I a_1^J$ we can modify these coefficients at will, since the variations read 
\equ{
 \gd_\ga \int a_1^I a_1^J = 
 \int \big( \d \ga^I a_1^J - \d \ga^J a_1^I \big) = 
 - \int \big( \ga^I f_2^J - \ga^J f_2^I \big)~. 
 \label{free_counter} 
}
The choice $c_{IJ} =1/2$ treats all mixed anomalies symmetrically.

As the interaction $\int a_1^I a_1^J$ can be supersymmetrized to $\int \d^2\gs \d^2\gth^+\, (V^I A^J - V^J A^I)$, the anomalous superpotential
\equ{ 
W_\text{anom}(\gTh) = 
\frac 1{2\gp}\, 
\Big( 
 \frac 12\, \sum_I \cA_{II} \, \gTh^I F^I  
+ \sum_{I<J} \cA_{IJ} \,
\Big[ (1 - c_{IJ} )\, \gTh^I F^J + c_{IJ}\, \gTh^J F^I \Big] 
\Big)~, 
\label{W_anom_c} 
}
is also only determined up to the coefficients $c_{IJ}$ for $I\neq J$. That is in \eqref{W_anom} we made the specific choice $c_{IJ} = 1/2$.

\subsection{Non--invariant Fayet--Iliopoulos terms}

A consistent GLSM has been obtained when all pure and mixed anomalies cancel. However, as observed by Adams et al.~\cite{Adams:2006kb,Adams:2009zg,Adams:2009tt} there is another possibility: One can involve a Green--Schwarz (GS) mechanism on the worldsheet. We first describe this mechanism in general and comment on its concrete realization later.

Consider the FI--superpotential \eqref{FI_term}. Let us assume that under a gauge transformation the FI--parameters $\gr_I$ are not invariant, but rather transform as 
\equ{
\gr_J \ra \gr_J +  \cT_{IJ}\, \gTh^I~, 
\label{Shift_trans} 
}
with $\cT_{IJ}$ some constants. These constants are in general not symmetric under the interchange of the gauge indices $I$ and $J$. (A single $\gr^J$ may be charged under various $U(1)$ gauge symmetries simultaneously.) Consequently, the FI--superpotential transforms as 
\equ{
W_\text{FI} \ra 
W_\text{FI} + \frac1{2\pi} \sum_I\, \cT_{II}\, \gTh^I\, F^I + 
\frac1{2\pi} \sum_{I<J} \,\Big( \cT_{IJ}\, \gTh^I\, F^J +\cT_{JI}\, \gTh^J\, F^I \Big)~. 
\label{FI_shift}
}
These transformations can be used to cancel the anomalous variation of the effective action $S_\text{eff}$ encoded in the anomalous superpotential \eqref{W_anom_c}, i.e.\ 
\(
\gd S_\text{eff} + \gd S_\text{FI} = 0. 
\)
The conditions to cancel the pure and mixed gauge anomalies read: 
\begin{subeqns} 
\equa{ 
\text{Pure anomalies:} \qquad & 
\cT_{II}  = \frac 12\, \Big( Q_I \cdot Q_I -  q_I \cdot q_I \Big)~, & I=J~,  
\label{anom_conditions_a}
\\[2ex] 
\text{Mixed anomalies:}  \qquad & 
\cT_{IJ} =   (1-c_{IJ})\, (Q_I \cdot Q_J -  q_I \cdot q_J)~,&I<J~, 
\label{anom_conditions_b}
\\[2ex] 
  & 
\cT_{JI} =   c_{IJ} \,( Q_I \cdot Q_J -  q_I \cdot q_J)~, & I>J~. 
\label{anom_conditions_c}
}
\end{subeqns}
Here we have included the freedom to shift mixed anomalies around by making specific choices for the coefficients $c_{IJ}$. As observed above, the GS--coefficients $\cT_{IJ}$ are often not symmetric, hence we need the $c_{IJ}$ freedom in order to increase the chance to cancel the anomalies.

\subsubsection*{Gauge vortex instantons}

Under many circumstances the GS--coefficients $\cT_{IJ}$ are subject to stringent quantization conditions which can be easily incompatible with the anomaly conditions \eqref{anom_conditions_a}--\eqref{anom_conditions_c}. To see how the quantization conditions on the GS--coefficients $\cT_{IJ}$ arise, we first recall some basic facts concerning gauge instantons in two dimensions \cite{Adams:2003zy,Adams:2009tt}: The BPS equations are
\equ{
\Big(\der_\gs + i \, \sum_I (q_I)^a\, a^I_{E\,\gs} \Big) z^a = 0~, 
\qquad 
- f^I_{E\,21} = \tD^I = \sum_a (q_I)^a\, |z^a|^2 - b^I~, 
}
in the Euclidean theory obtained after the Wick rotation: 
$\gs^0 = -i \gs^2$ so that $f^I_{\gs\bgs} = - i f^I_{E\,21}$. 
For a one--instanton solution the scalar $z^a$ vanishes at a single point on the worldsheet, say $\gs =0$, and the phase of $z^b$ winds non--trivially around this zero. Using polar coordinates $\gs(\gr, \gth) = \gr \exp(i\gth)$ on the Euclidean worldsheet we have asymptotically  \cite{Adams:2003zy}
\equ{ 
z^b(\gr,\gth) \sim \sqrt{\frac {b^J}{(q_J)^a}} \, e^{i \gth} +\ldots~, 
\qquad 
(q_J)^b a^J_{E\,1}(\gr,\gth) \sim \frac {\d \gth}{\gr} +\ldots~, 
}
for large $\gr$, where $+\ldots$ denote exponentially suppressed corrections, and $z^b(0) = a^J_{E\,1}(0) = 0$. The worldsheet gauge flux is then quantized as 
\equ{ 
\sum_J (q_J)^b\, \int \frac{ f^J_{E\,2}}{2\gp} = 1~,
\label{gauge_instanton} 
}
where $f^J_{E\,2} = f^J_{E\,21}\, \d \gs^2 \d \gs^1$. Consequently, for a set of $U(1)$ gauge multiplets we need to investigate non--trivial vortex solutions supported by all the charged chiral multiplets that only exist for specific values of the parameters $b^I$. In particular, when certain chiral superfields are charged under a multitude of $U(1)$ gauge symmetries, reading off the precise quantization of their gauge fluxes is rather involved. But if a chiral superfield is charged under a single $U(1)$ with charge $q$, it induces a quantization of the gauge flux in units of $1/q$. Hence a rough rule of thumb is that the charge with the largest absolute value sets the flux quantization unit.

\subsubsection*{Logarithmic Fayet--Iliopoulos terms}

With this in mind, we return to the quantization conditions on the coefficients $\cT_{IJ}$. We give an explicit construction of $\gr_J(\gPs)$ in the FI--action \eqref{FI_term} that transforms as specified in \eqref{FI_shift}. Given that the chiral multiplets $\gPs^a$ generically transform with chiral superfield phases, i.e.\ as given in \eqref{bos_gauging}, we can obtain $\gr_J$ that transform as shifts under gauge transformations. By taking 
\equ{ 
W_\text{log FI} = \frac 1{2\gp}\, \gr_J(\gPs) F^J~, 
\qquad 
\gr_J(\gPs) =  \gr^0_J +  T_{XJ} \, \ln  R^X(\gPs)~, 
\label{log_FI} 
}
we obtain the GS--coefficients 
\equ{
\cT_{IJ} = r_I^X  \, T_{XJ}~.
\label{modifiedAnomalyCancellation}  
}
Here $\gr^0_J$ are constants  and $R^X(\gPs)$ are homogeneous polynomials with $U(1)$ charges $r_I^X$.

The simplest choice for this would be to take $R^X(\gPs)$ equal to one of the chiral superfields $\gPs^a$, i.e.\ $r_I^X = q_I^a$. In this case the exponentiated Euclidean action $\exp(- S_\text{FI})$ is generically not invariant under a trivial phase redefinition $z^a \ra e^{2 \gp i} z^a$. As 
\equ{
\int \d^2 \gs\d\gth^+\, W_\text{FI} + \text{h.c.} \supset 
- i\, T_{aJ} \int \text{Im}( \ln z^a)\, \frac{f^J_{E\,2}}{2\gp}~, 
}
\(
e^{-S_\text{FI}} \ra e^{-S_\text{FI}} \;
\exp\big( 2 \gp i\, T_{aJ} \int \frac{f^J_{E\,2}}{2\gp} \big)
\) 
under such a trivial phase redefinition, and we obtain the quantization conditions 
\equ{ 
T_{aJ} \int \frac{f^J_{E\,2}}{2\gp}  \in  \Intr~. 
\label{Quant} 
}
Assuming that the chiral superfields with the largest absolute value of $U(1)$ charges are each charged under a single $U(1)$ only, this condition implies 
\(
T_{aJ} \in q_J\, \Intr, 
\)
in terms of the largest charges $q_J$.

Similar quantization conditions arise for more general logarithmic FI--terms \eqref{log_FI} with other homogeneous polynomials $R^X(\gPs)$. In this case the quantization condition involves the degree of this polynomial. For any choice the precise units of quantization of $T_{XJ}$ depend on the chosen $U(1)$ charge normalization. Of course, whether the anomaly conditions \eqref{anom_conditions_a}--\eqref{anom_conditions_c} can be solved in the presence of these quantization conditions does not depend on this normalization.  

\subsubsection*{Linear Fayet--Iliopoulos terms} 

In \cite{Adams:2006kb,Adams:2009zg,Adams:2009tt} a non--singular  covariant FI--term has been presented, that is linear in a chiral superfield $Y$: 
\equ{
W_\text{linear FI} = \frac 1{2\gp}\, m\, Y\, F~. 
\label{FI_linear} 
}
If this superfield transforms as a shift,  
\(
Y \ra Y + n\, \gTh,
\)
under a gauge transformation with parameter $\gTh$, we can use it to cancel some pure and mixed worldsheet gauge anomalies, i.e.\ the anomaly coefficient equals $\cT= mn$. The standard quadratic kinetic action has to be extended to 
\equ{
S_\text{kin} = 
\frac {R^2}4 \, \int \d^2\gs \d^2\gth^+\, 
\Big(Y + \bY + n\, V\Big) \Big(i\bder Y - i \bder \bY - n\, A\Big)~, 
}
to preserve gauge invariance. In order to ensure that the target space is compact,  $y = Y|$ has to be the complex coordinate on a flat torus 
\equ{
y \sim y + 1~, 
\qquad 
y \sim y + \gt~, 
}
where $\gt= \gt_1 + i\, \gt_2$ defines the complex structure of this torus. Its radius $R$ has been scaled out in front of the kinetic term. On an arbitrary instanton background the resulting action, containing  
\equ{
S_\text{kin} + S_\text{F} \subset
\int \d^2 \gs\, \Big\{ 
\Big(\frac m{2\gp} - \frac {n\, R^2}{2}\Big) \text{Re}(y)\, \tD + 
\Big(\frac m{2\gp} + \frac {n\, R^2}{2}\Big) \text{Im}(y)\, f_{\gs\bgs}  
\Big\}~,  
}
violates these torus periodicities, unless
\equ{
\gp\, R^2 =  \frac mn~, 
\qquad 
\frac 1{\gp}\, m \, \gt_2 
\int \frac{f_{E\,2}}{2\gp} \in \Intr~. 
}
Now, in order to cancel anomalies $m$ is fixed. This in turn leads to a quantization of the complex structure $\gt_2$ and the radius $R$ of this torus. The authors of refs.\ \cite{Adams:2006kb,Adams:2009zg,Adams:2009tt} use this to construct a GLSM realization of the torsional geometries \cite{Dasgupta:1999ss,Fu:2006vj,Fu:2008ga,Becker:2008rc}.

\subsection{Non--\Kh\ torsion geometry}

The inclusion of a non--constant FI--term \eqref{FI_term} leads to a target space geometry which is no longer \Kh~\cite{Adams:2006kb}: The Kalb--Ramond two--form $B_2$ can be expanded as $B_2(z) = \gb_I(z)\, F^I_2$ in harmonic two--forms $F^I_2$. Locally they are defined as $F^I_2 = \d A^I_1$ in the ambient space. The worldsheet gauge field one--forms $a^I_1 = a^I_\gs \d \gs + a^I_\bgs \d \bgs$ are, after imposing their equations of motion, the pull--backs of the connections $A^I_1$ on the target space.

The coefficients $\gb_I(z)$ can be interpreted as axions with a non--trivial background over the target space geometry. Given that the axions $\gb_I(z) = \text{Im}(\gr_I(z))$ transform with shifts under the worldsheet gauge transformations \eqref{Shift_trans}, the three--form field strength $H_3$ of $B_2$ has to be modified to \cite{Adams:2006kb}
\equ{
H_3 = \big( \d \gb_J +  r_I^X T_{XJ}\,  A_1^I \big) F_2^J~,
\label{H3_expansion} 
}
in order to be globally well--defined. This is the GLSM realization of the effect discussed in~\cite{Hull:1985jv,Hull:1986xn}: The anomalies in transformations of the worldsheet fermions induce the target space GS--mechanism.

Consequently, the target space is no longer \Kh, i.e.\ there is torsion, since $H_3 = i (\bder - \der) J_2 \neq 0$ implies that the fundamental two--form $J_2$ (i.e.\ \Kh--form for a \Kh\ manifold) is no longer closed. For this reason such GLSMs are sometimes called torsion GLSMs (TLSMs) \cite{Adams:2006kb}.

\subsubsection*{NS5 branes}

When the worldsheet FI--term becomes singular a more drastic modification of the target space geometry arises: NS5 branes appear. 
The Kalb--Ramond Bianchi identity is obtained by applying the exterior derivative $\d$ to \eqref{H3_expansion},
\equ{
\d H_3 = X_4 + \tr \cR_2^2 - \tr \cF_2^2~, 
\label{local_BI} 
}
where $\cR_2$ and $\cF_2$ are the anti--Hermitian target--space torsion--improved curvature and gauge field strengths, respectively. When $\gb_I$ can become singular, $\d (\d \gb_I) \neq 0$, there is an additional contribution
\(
X_4 = \d(\d \gb_J) F_2^J. 
\)
As $X_4$ measures the failure of $\tr \cR_2^2 - \tr \cF_2^2$ being exact, it signals the presence of NS5 branes \cite{Witten:1995gx,Tong:2002rq} (also sometimes referred to as H5 branes \cite{Strominger:1990et,Honecker:2006dt}). Even though the perturbative heterotic worldsheet theory is incapable of describing the microscopic dynamics of NS5 branes, it feels their effects.

Using the examples of non--invariant FI--terms we can make this more concrete: For example, for a logarithmic FI--term \eqref{log_FI}, with $\gr_J(\gPs) = \gr_J^0 + T_{aJ} \ln \gPs^a$,  
we find 
\equ{
\gb_J = T_{Ja}\, \ln \Big( \frac {z^a}{\bz_a} \Big)~, 
\qquad 
\d \gb_J = T_{Ja}\, \Big( \frac {\d z^a}{z^a} - \frac{\d \bz_a}{\bz_a} \Big)~, 
\qquad 
\d(\d \gb_J) = 2\gp\, T_{Ja}\, \gd^2(z^a)\, \d \bz_a \d z^a~. 
}
Since the internal profile of the axion $\gb_J$ is the two--form $F_2^J$, it follows that the NS5 brane is located at the intersection of the divisor $D_a := \{z^a = 0\}$ with the Poincar\'e dual of $F_2^J$. If $R^X(\Psi)$ is a polynomial rather than a monomial of a single superfield, one can interpret its coefficients as NS5 brane position moduli. However, since we only know the class of the divisor dual to $F_2^J$ but not the precise representative, the precise hypersurface it corresponds to is left unspecified. In particular when this class has more than one representative, the exact locus of the NS5 brane is not determined.

In certain phases of the GLSM the singularity of the logarithm can be avoided. For example, the scalar D--term potential on the worldsheet
\equ{
V_\text{D naive} = \frac {e_I^2}2 \, 
\Big( 
(q_I)^a \, |z^a|^2 + (q_I)^m \, |z^m|^2 
- \frac 1{2\gp} \, b_I^0 
\Big)^2~, 
\label{D_term_naive} 
}
might not allow a certain $z^a$ to vanish. When precisely this $z^a$ appears in the logarithmic FI--term, it will never become singular. In this phase we do not have any NS5 branes. But when we consider a transition to another phase, where the $z^a$ can become zero, NS5 branes are present. As certain phases of a GLSM may correspond to different coordinate patches, this means that the NS5 brane is only visible in certain patches but not in others. 

In the light of this interpretation that logarithmic FI--terms signal NS5 branes in the system, the coefficients $T_{XJ}$ can be viewed as counting the number of NS5 branes. The quantization condition \eqref{Quant} for a logarithm of a single chiral superfield precisely shows that $T_{aJ}$ are integers provided all worldsheet gauge instantons are supported by scalars of unit charge. However, often there are additional chiral superfields in the GLSM that have larger charges than the minimal one. Hence, unless there is a reason why they cannot induce gauge instantons, $T_{aJ}$ is quantized in units of the larger charges. In the NS5 brane interpretation one needs to have a set of NS5 branes. In the examples discussed below we return to this issue in a concrete setting in order to understand the potential consequences.

\subsubsection*{NS5 brane backreaction}

So far the analysis of the resulting geometry of the GLSM has been performed without considering possible geometrical backreactions. We now consider this important effect: In the presence of the logarithmic FI--term \eqref{log_FI} the naive worldsheet D--term potential \eqref{D_term_naive} gets corrected to 
\equ{
V_\text{D} = \frac {e_I^2}2 \, 
\Big( 
(q_I)^a \, |z^a|^2 + (q_I)^m \, |z^m|^2 
- \frac 1{2\gp}(b_I^0 +T_{XI} \, \ln|R^X(z)|)
\Big)^2~. 
}
In particular, when $R^X(\gPs) = \gPs^a$, this implies that it is not possible anymore to set $z^a=0$. This means that the corresponding divisor $D_a := \{z^a = 0\}$ no longer exists: The infinitely thin NS5 brane is replaced by a non--trivial modification of the target space geometry near the position where this brane used to be.

\subsection{Orbifold modular invariance versus Bianchi identities}

The inclusion of the field--dependent FI--terms might help to resolve the following longstanding paradox between orbifold and CY model building: The basic consistency conditions of a $\Intr_N$ orbifold theory are the modular invariance conditions 
\equ{
\frac N2 \Big( V_r^2 - v_r^2 \Big)  = 0~\text{mod}~1~,  
\label{mod_inv} 
}
for the local orbifold twist $v_r$ and shifts $V_r$, where $r$ labels the different orbifold fixed points. Note that these are not strict but only mod conditions. In contrast, the standard conditions for CY compactifications, the integrated Bianchi identities over any closed four--cycle $C$,
\equ{
\int_C \Big(\tr \cF_2^2 - \tr \cR_2^2 \Big)  = 0~,
\label{int_BI} 
}
are strict (i.e.\ not mod) conditions. In this sense the Bianchi identities \eqref{int_BI} on a CY are much stronger conditions than the modular invariance conditions \eqref{mod_inv} on its orbifold counter part. However, from the first set of  conditions in \eqref{anom_conditions_a}--\eqref{anom_conditions_c} we see that the GS mechanism on the worldsheet precisely weakens the pure gauge anomaly conditions, which may be identified with the integrated Bianchi identities, to identical mod conditions of the orbifold theory: The GLSM corresponding to a resolution of an orbifold model has charges quantized in units of $1/N$. Hence by multiplying the pure anomaly cancellation conditions by a factor $N$ they have the same structure as the modular invariance conditions \eqref{mod_inv}, and the right--hand--side is indeed integral.

Once the orbifold modular invariance conditions are fulfilled, adding an arbitrary big $E_8 \times E_8$ lattice vector to the orbifold shift vectors and Wilson lines cannot spoil the relations anymore. Adding such lattice vectors generically makes the $\sum Q^2$ larger, which can be compensated by NS5 branes, presumable without breaking supersymmetry.

\section{Examples} 

\subsection{Logarithmic FI--terms in the GLSM of the quintic}
\label{sc:Quintic} 

In this example we consider the familiar quintic $\mathbbm{P}^4[5]$. This CY is obtained as a hypersurface defined by a homogeneous degree five polynomial $P(\gPs)$ in $\mathbbm{P}^4$. The GLSM that describes this model has the following charge assignment: 
\equ{
\arry{|c||c | c | c | c |}{
\hline 
\text{superfield} & \gPs^{a=1,\ldots, 5} & \gG & \gL^{\ga=1,\ldots n} & \gF
\\ \hline 
\text{lowest component} & z^a & \gg & \gl^\ga & x 
\\ \hline \hline 
\text{gauge charge} & 1 & -5 & Q^\ga & -5 
\\ \hline 
}
}
The total number $n$ of charged chiral--Fermi multiplets $\gL^\ga$ as well as their charges $Q^\ga$ are left arbitrary in the following discussion. The geometry is determined by the superpotential 
\equ{
W_\text{geom} = \gG \,P(\gPs)~, 
\quad\text{e.g.}\quad 
P(\gPs) = \sum_a (\gPs^a)^5~, 
\label{GeometryEqnQuintic}
}
and the D--term potential 
\equ{
V_\text{D naive} = \frac 12 \Big( \sum_a |z^a|^2 - 5\, |x|^2 - b \Big)^2~. 
}
The gauge bundle is constrained by 
\equ{
W_\text{bundle} = \gF\, N_\ga(\gPs)\, \gL^\ga~, 
\label{bundle_SP_Quintic} 
}
where $N_\ga(\gPs)$ are polynomials in $\gPs^a$ of degree $5-Q^\ga$ that should not all vanish at the same time. As a consequence, we see that the F--term equations of $\gL^\ga$ in the geometrical regime, where at least one of the $z^a$ takes a VEV, forces the VEV of $x$ to vanish identically.

Let us we investigate which kind of anomalies can be cancelled upon inclusion of a logarithmic FI--term \eqref{log_FI}. For the sake of simplicity we consider one particular choice for the homogeneous polynomial $R^X$: 
\equ{
W_\text{log\ FI} = \frac 1{2\gp} \Big( \gr^0 + T \ln [R^X(\gPs)]\Big)\, F~, 
\qquad 
R^X(\gPs) = \sum\limits_{a=1}^5 c_a \gPs^a~,
\label{logarithmicCounterTermQuintic}
}
with $c_a$ constants. For this choice $r^X=1$, so the relation \eqref{modifiedAnomalyCancellation} between the FI--parameter $T$ and the anomaly coefficient $\cT$ gives the anomaly equation 
\eqref{anom_conditions_a}:
\equ{
T =  \mathcal{T} =  \frac12 \sum_\ga (Q^\ga)^2 - \frac 52~.
\label{Anomalies_Quintic}
}
To understand how restrictive this anomaly condition is, we investigate the quantization of the anomaly coefficient $T$.  First of all we notice that in general in this theory two types of gauge instantons are possible: i) gauge instantons, where one of the $\gPs^a$ is non--zero, have a minimal non--zero flux: $\int f_{E\,2}/(2\gp) = 1/1 = 1$;  ii) gauge instantons, where $\gF$ is non--zero, have: $\int f_{E\,2}/(2\gp) = 1/(-5) = -1/5$. Consequently, equation \eqref{Quant} implies that $T\in 5\, \Intr$ in the above normalization where the minimal gauge charge is $1$ and the largest $-5$.

Next we analyze what happens when $T$ takes specific values. If $T=0$, the logarithmic term does not occur and the anomalies cancel exactly. The other cases $T>0$ and $T<0$ are more interesting for us:

\subsubsection*{$\mathbf{T<0:}$}

In this case no solutions to the anomaly condition \eqref{Anomalies_Quintic} can be found.  The quantization condition requires $T$ to be a multiple of five, while the right--hand--side of \eqref{Anomalies_Quintic} is bounded from below by $-5/2$. For negative $T$ these requirements are incompatible. Note that this is due to the strong quantization condition which says that $T\in 5 \, \Intr$. If the condition was that $T$ was just an integer, solutions would be possible.

\subsubsection*{$\mathbf{T>0:}$} 

In this case $c_2(TX) > c_2(V)$, since $\sum_\ga Q_\ga^2 > \sum_a q_a^2$, and therefore NS5 branes can in principle compensate their mismatch.  A concrete realization is given by $Q = (1^{15})$, which gives the minimal non--trivial $T = 5$ solution to the quantization condition with NS5 branes.

The field dependent FI--term modifies the D--term potential to 
\equ{
V_\text{D} = 
 \frac 12 \Big( \sum\limits_{a=1}^5 |z^a|^2 - 5\, |x|^2 - \frac 1{2\gp}(b 
 + T\, \ln |z^1|)  \Big)^2~. 
 \label{V_D_Quintic} 
}
Here we have assumed that only $\gPs^1$ appears under the logarithm in \eqref{logarithmicCounterTermQuintic}, i.e.\ $c_1=1$, $c_a=0$ for $a=2,\ldots,5$. Hence this potential clearly forbids $z^1=0$; the geometrical backreaction has shielded the NS5 brane.

\subsection{NS5 branes on the Calabi--Yau $\mathbf{\mathbbm{P}^7[2,2,2,2]}$ }
\label{sc:P7} 

Both of our next examples deal with the CY threefold  ${\mathbbm{P}^7[2,2,2,2]}$ which is obtained as a complete intersection of four degree--two hypersurfaces in the seven dimensional projective space $\mathbbm{P}^7$.

We first consider an $SU(6)$ gauge bundle on this geometry. The GLSM describing this model has the following charge assignment: 
\equ{
\arry{|c||c | c | c | c |}{
\hline 
\text{superfield} & \gPs^{a=1,\ldots, 8} & \gG^{\gm = 1,\ldots,4} & \gL^{\ga=1,\ldots 8} & \gF^{m=1,2}
\\ \hline 
\text{lowest component} & z^a & \gg^\gm & \gl^\ga & x^m 
\\ \hline \hline 
\text{gauge charge} &  1 & -2 & 1 & -4
\\ \hline 
}
\label{charges_P7}
}
The geometry and the gauge bundle are encoded in the superpotentials  
\equ{
W_\text{geom} = \gG^{\gm} \,P_{\gm}(\gPs)~, 
\qquad 
W_\text{bundle} = \gF^m \, N_{m\ga}(\gPs)\, \gL^\ga~, 
\label{bundle_SP_Exmpl} 
}
where $P_{\gm}(\gPs)$ and  $N_{m\ga}(\gPs)$ are generic polynomials of degree two and three, respectively.

We can write down the following logarithmic FI--term 
\equ{
W_\text{log FI} = \frac 1{2\gp} \Big( \gr^0 + T \ln \Big[ \sum_a c_a \gPs^a\Big] \Big)\, F~. 
\label{FI_P7}
}
The FI--parameter $T$ is again quantized: The most stringent conditions come from the instantons involving the scalars with the largest absolute value of the $U(1)$ charge, which are the $x^m$. Consequently,  condition \eqref{Quant} implies that $T \in 4\mathbbm{Z}$ for this model. With the charge assignment given in \eqref{charges_P7} the worldsheet GS anomaly cancellation condition 
\equ{
T =  \cT 
= \frac 12 \Big( 
8 \cdot (1)^2 + 4 \cdot (-2)^2 - 8 \cdot (1)^2 - 2 \cdot (-4)^2 
\Big) 
= -8~, 
\label{Anomalies_Exmpl} 
}
is compatible with this quantization condition with negative $T$, which indicates that there anti--NS5 branes in the system.

In the presence of the logarithmic FI--term \eqref{FI_P7}, the D--term potential reads 
\equ{
V_\text{D} = 
\frac 12 \Big( \sum_a |z^a|^2 - 4 \sum_m |x^m|^2 - \frac 1{2\gp} (b 
-|T|\, \ln |z^1|)  \Big)^2~, 
\label{V_D_Exmpl} 
}
where we again assumed that only $\gPs^1$ is involved in the logarithmic FI--term, i.e.\ $c_a = \delta_{1a}$ in \eqref{FI_P7}. Furthermore, we have made explicit that $T=-|T|$ is negative. In the geometrical regime where $x=0$, the D--term potential leads to the constraint 
\equ{
  \sum_{a=2}^8 |z^a|^2  =  \frac 1{2\gp}(b - |T|\, \ln |z^1|)  -  |z^1|^2~.
}
This equation has far reaching consequences for the geometry: Even though the original CICY is compact, the geometry described by the model with $T<0$ is not. We can make the $z^a$ on the left--hand--side arbitrary large by taking $z^1$ closer and closer to zero. This effect we may take as evidence that the system contains anti--NS5 branes.

It is also possible to construct a vector bundle for this CICY which leads to NS5 branes instead of anti--branes. This model has the GLSM charge assignment: 
\equ{
\arry{|c||c | c | c | c |}{
\hline 
\text{superfield} & \gPs^{a=1,\ldots, 8} & \gG^{\gm = 1,\ldots,4} & \gL^{\ga=1,\ldots 4} & \gF^{m=1,2}
\\ \hline 
\text{lowest component} & z^a & \gg^\gm & \gl^\ga & x^m 
\\ \hline \hline 
\text{gauge charge} &  1 & -2 & 1 & -2
\\ \hline 
}
\label{charges_P7prime}
}
Therefore this GLSM describes an $SU(2)$ bundle. In this case we find for the anomaly: 
\equ{
T = \cT = 
\frac 12 \Big( 
4 \cdot (1)^2 + 4\cdot (-2)^2 - 8 \cdot (1)^2 - 2 \cdot (-2)^2
\Big) = 2~, 
}
which is positive and hence indicates that NS5 branes are present. Moreover, since now the largest absolute value of a chiral superfield charge is 2, the quantization condition is met in the minimal possible way.

In this case the D--term potential reads 
\equ{
V_\text{D} = 
\frac 12 \Big( \sum_a |z^a|^2 - 2 \sum_m |x^m|^2 - \frac 1{2\gp} (b 
+ T\, \ln |z^1|)  \Big)^2~, 
\label{V_D_Exmplprime} 
}
where we have again assumed that only $\gPs^1$ is involved in the logarithmic FI--term, i.e.\ $c_a = \delta_{1a}$ in \eqref{FI_P7}. 
Hence we are able to keep the model compact in the presence of gauge anomalies, since $T=2$ is positive.

Finally, since in this setting the Bianchi identity defines an effective class,
\equ{
\text{ch}_2(V) - \text{ch}_2(TX) = 
\frac12 \left( \tr \cF_2^2 -  \tr \cR_2^2 \right)  =  2 D_1 D'~, 
}
we interpret the resulting geometry as having NS5 branes wrapping an effective curve in the class $D_1 D'$. Here $D_1 = \{ z^1 = 0 \}$ and $D'$ is a divisor associated to the restriction of the hyperplane class of $\mathbbm{P}^7$ onto the CICY. This sheds some light onto the meaning of the coefficients $c_a$ in the polynomial under the logarithm in \eqref{FI_P7}: They can be interpreted as NS5 brane position moduli. The choice of $c_1=1, c_{a\neq1}=0$ positions the NS5 brane onto a curve which is contained in $D_1$. As we observed in general, since from this equation we can only infer the class of $D'$ but not its exact representative, it does not completely pinpoint the position of the NS5 brane.

\subsection{Further examples}

The examples studied in subsection \ref{sc:P7} all correspond to stable or polystable bundles\footnote{We thank James Gray for helping checking.} in the smooth geometrical phase. In the paper \cite{Anderson:2007nc} a list of stable monad bundles $V$ on the five possible CICYs $X$ in a single projective space are given for which the index is a multiple of three. As the authors indicate these models require $T = c_2(X) - c_2(V)$ NS5 branes. The quantization condition on the FI--coefficient $T$ found in our work seems to indicate that these models only admit a consistent GLSM description when $T$ is divisible by all number $c_i$ (the charges of the chiral superfield $\Phi$ in our language).

A further example to which our results could be applied concerns the resolved conifold with torsion and NS5 branes \cite{Carlevaro:2009jx}. Their construction seems to be similar to our example in section \ref{sc:Quintic}: Making the sum of charges of the Fermi superfields large can always be compensated by a set of NS5 branes.

\bibliographystyle{paper}
{\small
\providecommand{\href}[2]{#2}\begingroup\raggedright\endgroup

}

\end{document}